\documentclass[twocolumn,preprintnumbers,amsmath,amssymb,superscriptaddress]{revtex4}
\usepackage{graphicx}
\usepackage{dcolumn}
\usepackage{bm}
\usepackage{float}
\usepackage{afterpage}

\begin{document}

\title{Magnetic field dependent interactions in an ultracold Li-Yb($^3$P$_2$) mixture}
\author{William Dowd}
\affiliation{Department of Physics, University of Washington, Seattle WA 98195, USA}
\author{Richard J. Roy}
\affiliation{Department of Physics, University of Washington, Seattle WA 98195, USA}
\author{Rajendra K. Shrestha}
\affiliation{Department of Physics, University of Washington, Seattle WA 98195, USA}
\author{Alexander Petrov}
\altaffiliation{Alternative address: St. Petersburg Nuclear Physics Institute, Gatchina, 188300; Division of Quantum Mechanics, St. Petersburg State
University, 198904, Russia.}
\affiliation{Department of Physics, Temple University, Philadelphia PA 19122, USA}
\author{Constantinos Makrides}
\affiliation{Department of Physics, Temple University, Philadelphia PA 19122, USA}
\author{Svetlana Kotochigova}
\affiliation{Department of Physics, Temple University, Philadelphia PA 19122, USA}
\author{Subhadeep Gupta}
\affiliation{Department of Physics, University of Washington, Seattle WA 98195, USA}
\date{\today}
\begin{abstract}
Magnetic Feshbach resonances have allowed great success in the production of ultracold diatomic molecules from bi-alkali mixtures, but have so far eluded observation in mixtures of alkali and alkaline-earth-like atoms. Inelastic collisional properties of ultracold atomic systems exhibit resonant behavior in the vicinity of such resonances, providing a detection signature. We study magnetic field dependent inelastic effects via atom loss spectroscopy in an ultracold heteronuclear mixture of alkali $^6$Li in the ground state and alkaline-earth-like $^{174}$Yb in an excited electronic metastable state ($^3P_2$, $m_J = -1$). We observe a variation of the interspecies inelastic two-body rate coefficient by nearly one order of magnitude over a $100-520\,$G magnetic field range. By comparing to ab-initio calculations we link our observations to interspecies Feshbach resonances arising from anisotropic interactions in this novel collisional system.
\end{abstract}

\maketitle

\section{Introduction}

The ground electronic doublet-sigma ($^2\Sigma$) state of a diatomic molecule composed of an alkali and an alkaline-earth-like atom is endowed with an unpaired electron. This electronic degree of freedom distinguishes it from the spinless singlet-sigma ($^1\Sigma$) ground state of the bi-alkali molecules familiar to the ultracold gas community. In particular, it opens up new possibilities in the field of quantum simulation that exploit both electric and magnetic dipole moments \cite{SIM1}, and provides new opportunities to test fundamental physical theories through, for instance, electron electric dipole moment measurements \cite{FUND1}.

Magnetic Feshbach resonances\cite{FESH1,FESH2,FESH3} have been crucial in the production of ground-state bi-alkali molecules \cite{Cs2, KRb}, allowing for the transformation of free atoms into weakly-bound ``Feshbach" molecules. These Feshbach molecules can then be transferred into the ground state using coherent two-photon processes. Current experimental efforts with ultracold mixtures of alkali and alkaline-earth-like atoms \cite{YBLI1,YBLI2,RBYB1,RBSR1} now face the challenge of locating and utilizing suitable Feshbach resonances to create doublet-sigma molecules. Unfortunately, recent theoretical studies indicate that Feshbach resonances between the ground states of alkali and alkaline-earth-like atoms are extremely narrow and located at inconveniently high fields \cite{NEWFESH}. Such resonances have not yet been experimentally observed.

On the other hand, alkaline-earth-like atoms offer an additional avenue of collisional physics to explore due to the presence of long-lived (metastable) electronically excited $^3P$ states. Earlier studies of these metastable states in alkaline-earth atoms calcium and strontium \cite{CALC1,SR1} and alkaline-earth-like ytterbium \cite{Yb1} revealed both elastic and inelastic collisional behavior. Resonances between ground and metastable ytterbium were recently reported\cite{YBMETAFESH} and attributed to anisotropy in the atomic interactions\cite{METAFESH}. Our group recently realized a heteronuclear mixture of metastable ytterbium and alkali lithium \cite{LIYBMETA}, where broad interspecies Feshbach resonances of several Gauss width are theoretically predicted \cite{LIYBMETAFESH}. Such resonances would be accompanied by peaks in the interspecies inelastic collision rate.

\begin{figure}
\includegraphics[width=3.3in]{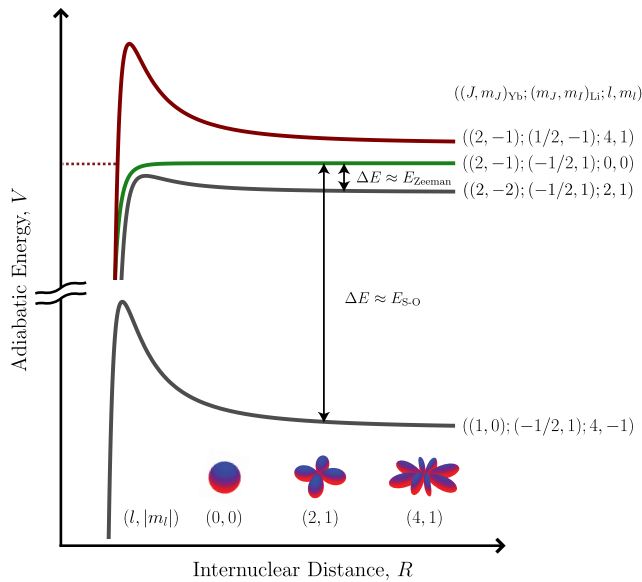}
\caption{(Color online). Schematic of Yb*+Li two-body collisions showing a subset of the various possible adiabatic potentials at a finite magnetic field.
Here we choose a coordinate system with projection quantum numbers defined along the external magnetic field direction. The channels are represented by the
quantum numbers: $((J,m_J)_{\rm Yb};(m_J,m_I)_{\rm Li};l,m_l)$, where $(J,m_J)_{\rm Yb}$ give the total angular momentum and projection of the metastable Yb* atom,
$(m_J,m_I)_{\rm Li}$ give the projections of the electronic and nuclear angular momenta for the Li atom, and $l$ is the relative nuclear orbital angular momentum
with projection $m_l$. The collision starts from the $((J,m_J)_{\rm Yb}=(2,-1);(m_J,m_I)_{\rm Li}=(-1/2,1);l=0,m_l=0)$, $s-$wave channel (green). In this
coordinate system the rotational and Zeeman interactions as well as the isotropic dispersion potential shift molecular levels, whereas the magnetic dipole-dipole
and anisotropic dispersion interactions lead to coupling between different rotational (with $\Delta l$ even) and Zeeman components. In particular, Feshbach
resonances occur in channels with a dissociation energy above the entrance channel (red), whereas the inelastic losses are due to a coupling to the energetically
lower channels (grey).}
\end{figure}

In this paper we report on the collisional properties of $^{174}$Yb in the metastable $^3P_2 (m_J=-1)$ state both with itself, and with ground state $^6$Li over a wide range of magnetic fields. We observe nearly one order of magnitude variation in the interspecies inelastic two-body collision rate coefficient. We also perform ab-initio calculations for this collisional system which link our experimental observations to the different collision channels (see Fig. 1) and to the presence of anisotropy-induced interspecies Feshbach resonances.

\section{Preparation of the ultracold Li-Yb* Mixture}

The experiment is performed in our mixture apparatus described in reference \cite{PRODUCTION}. We sequentially load compressed magneto-optical traps of Yb and Li into a single beam $1064\,$nm optical dipole trap. Orthogonal to this main beam, we add a much weaker beam ($1070\,$nm) which serves to increase the axial confinement. After $4\,$s of forced evaporative cooling consisting of lowering the intensity of the main beam, and concurrent interspecies thermalization, we obtain a mixture with $500(25)\times10^3$ Yb(Li) atoms at 1.1$\mu$K, with Yb in the ground ${^1S_0}$ state and Li in the ground ${^2S_{1/2}}$ state, spin-purified to the lowest energy Zeeman state ($|m_J,m_I\rangle=|-1/2,+1\rangle$ at high field).

We achieve spin-purified metastable ytterbium in the $^3P_2$, $m_J\,=\,-1$ state (Yb*) using the scheme detailed in prior work \cite{LIYBMETA}. Briefly, we optically pump using the ${^1S_0} \rightarrow {^3D_2}$ electric quadrupole transition at $404\,$nm\cite{BOWERS} which decays to the ${^3P_2}$ state with a 12\% branching ratio. The remaining fraction decays to ${^3P_1}$ which returns to ${^1S_0}$ with a $<1\,\mu$s decay time and the cycle repeats. We use $1\,$mW of $404\,$nm light focused to $30\,\mu$m aligned along the long axis of our dipole trap. In $10\,$ms we achieve 25\% transfer to the metastable state, populating the $m_J = -1$ and -2 sublevels. The ${^3P_2}$ substate polarizabilities are sensitive to the angle between the electric field of the trapping laser and the magnetic quantization axis. We exploit this effect to maximize (minimize) the trap depth for the $m_J = -1(-2)$ state by orienting the linear polarization of our optical trap parallel to the magnetic bias field (vertical in our case). Together with a $50\,$G/cm vertical magnetic field gradient applied against gravity, this spills the $m_J = -2$ atoms, resulting in a pure gas of Yb* atoms \cite{footspinpure}. Any atoms remaining in the ground state are subsequently removed by a resonant laser pulse on the strong ${^1S_0} \rightarrow {^1P_1}$ transition. At the end of the transfer process there are 2(1)$\times10^4$ Yb*(Li) atoms at $1.7(1.5)\,\mu$K with a peak density of $1.0(5.1)\times10^{12}/$cm$^3$ for Yb*(Li).

The entire transfer process is done at a moderate field of $100\,$G which is a compromise between minimizing losses due to inelastics during the transfer process and ramp time to the largest magnetic fields. Once transfer is complete, we ramp the electromagnets at $50\,$G/ms to the desired field. Coincident with this, we lower the magnetic gradient to $20\,$G/cm, eliminating the effect of gravitational sag for the Yb* atoms and hence maximizing the spatial overlap with the Li atoms. We allow the gases to interact for up to $40\,$ms before imaging the remaining atoms.

\begin{figure}
\includegraphics[width=3in]{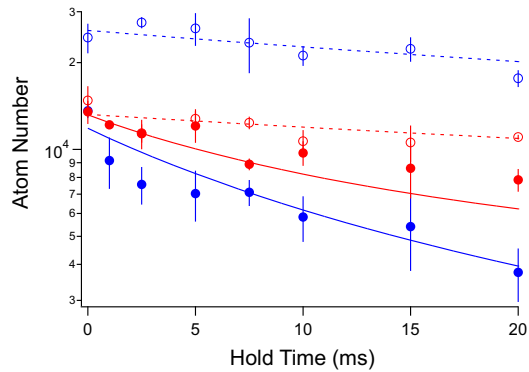}
\caption{(Color online). Sample lifetime curves of Yb*(blue) and Li(red) taken at $100\,$G (open circles) and $450\,$G (closed circles) are shown. Also shown are best fit curves (see text) for both $100\,$G (dashed) and $450\,$G (solid) datasets. All error bars are statistical.}
\end{figure}

\section{Analysis of field-dependent interactions}

Since the temperature of the system is far below the $p$-wave threshold for the system \cite{LIYBMETA}, we consider only $l=0$ rotational states in the entrance channel for collisions between $m_{\rm Yb}=m_{J,{\rm Yb}}=-1$ and $m_{\rm Li}=m_{J,{\rm Li}}+m_{I,{\rm Li}}=+1/2$ atoms. We analyze the time evolution of our 
ultracold mixture at different magnetic fields $B$ in terms of two field-dependent two-body inelastic parameters representing all Yb*-Yb* and all Yb*-Li inelastic 
collisions.

The time evolution of the atomic densities in our two-species mixture (Fig. 2) is fit according to the coupled differential equations
\begin{equation}
\dot n_{\rm Yb}=-K_2'(B)n_{\rm Yb}n_{\rm Li}-2K_2(B)n_{\rm Yb}^2
\end{equation}
\begin{equation}
\dot n_{\rm Li}=-K_2'(B)n_{\rm Yb}n_{\rm Li}
\end{equation}
where $n_{\rm Yb(Li)}$ is the Yb*(Li) density and $K_2(K_2')$ are the field dependent Yb*+Yb*(Yb*+Li) inelastic coefficients. Decay rates for single-body processes and for Li-Li interactions are negligible for this system \cite{LIYBMETA} and are therefore excluded from this analysis. Three-body decay rates of the form $K_3n_{\rm Yb}^3$ and $K_3'n_{\rm Yb}^2n_{\rm Li}$ can also contribute to system dynamics, but are expected to be much weaker given the relatively low densities involved.

The extracted inelastic coefficients depend sensitively on the spatial overlap of the gases, which is affected by temperatures, trap frequencies, and spatial offsets. The temperature itself is a dynamic quantity; atoms from the densest (and thus lowest energy) regions of the trap are preferentially removed, thus increasing the temperature. In addition, the excitation scheme preferentially heats the Yb* atoms from photon recoil. To guard against this, we measure the temperature dynamics of the two gases at several magnetic fields, and find that the temperature does not appreciably change over the short timescale of the experiment \cite{foottemp}.

We pay careful attention to the spatial overlap of the two clouds during the evolution time, limiting our analysis to the duration within which the two species have not lost more than 10\% overlap. This typically limits the datasets for analysis to the $0-20\,$ms regime \cite{footoverlap}.

At each magnetic field, we take a separate set of data with all Li atoms removed from the trap. This allows us to independently fit the $K_2$ parameter associated with Yb*+Yb* inelastic processes. This results in dramatically improved fits and uncertainties for the $K_2'$ parameter.

\begin{figure}
\includegraphics[width=3in]{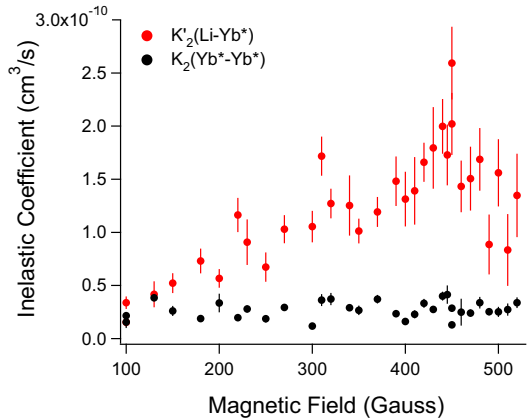}
\caption{(Color online). Spectra of inelastic decay coefficients as a function of magnetic field for Yb*+Yb* (black) and Yb*+Li (red). All error bars are statistical.}
\end{figure}

The resulting spectrum of inelastic rates for $100-520\,$G is shown in Fig. 3. We find that the intra-species inelastic rate is much weaker than the inter-species rate. The intra-species coefficient $K_2(B)$ remains mostly constant throughout this range. The inter-species coefficient $K_2'(B)$, however, displays an overall growth with magnetic field from $100-450\,$G of about a factor of eight. Of particular note is the peak at $450\,$G which approaches the unitarity limit of $2.9\times10^{-10}\,$cm$^3$/s \cite{JulienneUnitarity} for two-body Li-Yb interactions at these temperatures.

\begin{figure}
\includegraphics[width=3in]{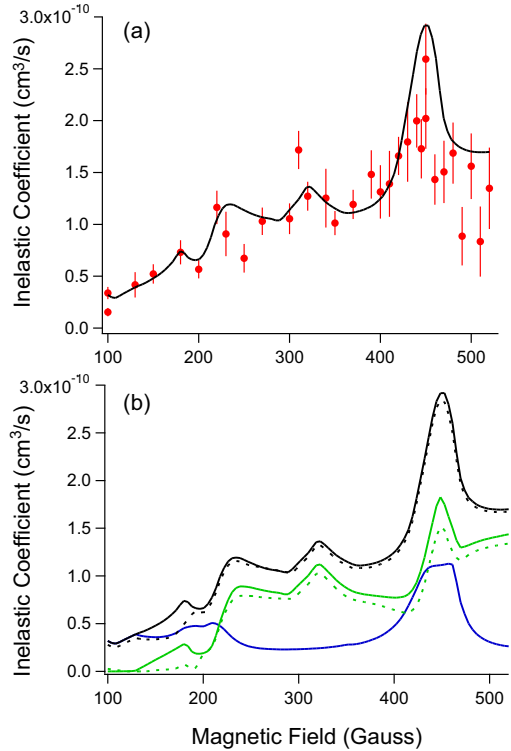}
\caption{(Color online). (a) shows the theoretical (solid black) curve with optimized short-range potentials (see text) for the inelastic loss rate coefficient between $^6$Li and $^{174}$Yb together with the experimental data (red points). (b) Inelastic loss rate to the ${^3P_2} m_J\,=\,-2$ (green) and ${^3P_1}$ (blue) channels together with the total loss rate (black). The dashed lines correspond to the same calculations performed without $l=0$ channels.}
\end{figure}

\section{Theoretical Model}


Our first-principles coupled-channel model is similar to the anisotropic scattering of highly-magnetic lanthanide atoms \cite{METAFESH}, and allows us to calculate both elastic and inelastic rate coefficients between colliding fermionic $^6$Li and bosonic metastable $^{174}$Yb atoms. The model treats the effects of the atomic energy splittings, molecular rotation, the magnetic dipole-dipole, and electrostatic isotropic and anisotropic dispersion interactions non-pertubatively and on equal footing. For bosonic $^{174}$Yb, the metastable state has a large spin-orbit interaction. Its nuclear spin is zero, thus there is no nuclear hyperfine structure. The fermionic $^6$Li has non-zero nuclear spin ${\vec \imath}_{\rm Li}$ and therefore both Zeeman and hyperfine splittings are present.

We perform close-coupling calculations in the atomic basis $Y_{lm_l}(\theta,\phi) |(i_{\rm Li}s_{\rm Li})Fm_{F}(s_{\rm Yb}l_{\rm Yb})Jm_{J} \rangle$, where $Y_{l m_l}(\theta,\phi)$ is a spherical harmonic and the angles $\theta$ and $\phi$ give the orientation of the internuclear axis relative to the magnetic field direction. All projections are defined relative to the magnetic field direction and we have used $l_{\rm Li}=0$ and $i_{\rm Yb}=0$. In this basis the hyperfine $a_{\rm hf}({\vec\imath}_{\rm Li} \cdot {\vec s}_{\rm Li})$, spin-orbit $a_{\rm so}({\vec s}_{\rm Yb} \cdot {\vec l}_{\rm Yb})$, and rotational interactions $\hbar^2 {\vec l}^2 /(2\mu_rR^2)$ are diagonal, where $\mu_r$ is the reduced mass of the molecule.

The total projection of the colliding $^6$Li($^2$S$_{1/2}$) and $^{174}$Yb(${^3P_2}$) is $M=m_{\rm Li}+m_{\rm Yb} +m_l= -1/2$, and is conserved. Inelastic 
atom-atom processes are present due to spin-orbit relaxation to the energetically-lower Zeeman sub-levels and $^3P_{1,0}$ multiplets.

The $R$-dependent coupling matrix elements between these atomic channels are due to molecular interactions and are evaluated using a small set of spherical tensor operators. There are three contributions: (i) a spin-independent isotropic potential $V_{\rm iso}(R)$, which is proportional to $1/R^6$ for large separation, (ii) a short-range isotropic exchange interaction described by the tensor $V^{\Sigma,\Pi}_{\rm ex}(R)[{\vec s}_{\rm Li} \otimes {\vec s}_{\rm Yb} ]_{00}$, which splits the doublets from the quartets of the $\Sigma$ and $\Pi$ potentials, where $V_{\rm ex}(R)$ falls off exponentially for large $R$, and (iii) an anisotropic 
quadrupole-like interaction $V_{\rm ani}(R) [\hat{C}_{2} \otimes [{\vec l}_{\rm Yb} \otimes {\vec l}_{\rm Yb}]_{2}]_{00}$, which lifts the degeneracy of the $\Sigma$ and 
$\Pi$ potentials, where $V_{\rm ani}(R)\propto 1/R^6$ for large $R$, and ${\hat C}_{kq}(\theta,\phi)=\sqrt{4\pi/(2k+1)} Y_{kq}(\theta,\phi)$. The 
interaction strengths $V_{\rm iso}(R)$, $V_{\rm ex}(R)$, and $V_{\rm ani}(R)$ are constructed such that, in the body-fixed frame with projections along the 
internuclear axis, the sum of four interactions reproduces our four non-relativistic potentials at short range and the non-relativistic van der Waals coefficients at 
long range \cite{SvetTheory}. In addition, we include two magnetic dipole-dipole interactions $c_1 [{\hat C}_2 \otimes [{\vec s}_{\rm Li} \otimes {\vec s}_{\rm Yb}]_{2}]_{00}/R^3$ and $c_2 [ {\hat C}_2 \otimes[{\vec s}_{\rm Li} \otimes {\vec l}_{\rm Yb}]_2]_{00}/R^3$, where $c_1 = -6.53\times 10^{-5} E_ha_0^3$ and $c_2 = -1.31\times 10^{-4} E_ha_0^3$.

The resultant inelastic scattering coefficient for fields between $100$ Gauss and $520$ Gauss is shown together with the experimental values in Fig.~4(a). We note that the agreement between our experimental and theoretical spectrum is only obtained after a study of the dependencies of the rate coefficient on the short-range shape of the $^{2,4}\Sigma^+$ and $^{2,4}\Pi$ electronic potentials. We cannot exclude the existence of other shapes of potentials that will lead
to loss rates consistent with the experimental data. 

We observe one clear resonance at $450\,$G, and some weak modulations at lower fields. These resonances have multi-channel dependence and cannot be characterized by a single partial wave, only converging to their final locations when collision channels up to $l=8$ are included in the calculation (see Fig.~4(b)). Our bound-state calculation of the LiYb* molecule allows us to assign the $450\,$G feature as arising from a resonance with a bound state of 60\% $g$-wave and 40\% $d$-wave character, and $m_l=1$. Other quantum labels for this resonance arise from the eigenstates of the atomic Zeeman plus hyperfine Hamiltonian \cite{SvetTheory} of $^6$Li and $^{174}$Yb, with $(m_J,m_I)_{\rm Li}=(1/2,-1)$ and $(J,m_J)_{\rm Yb}=(2,-1)$ for ytterbium. We find that the dominant loss process is to the Yb ${^3P_2}, m_J = -2$ manifold (see Fig.~4(b)), suggesting that the inelastic rate can be reduced significantly by eliminating this decay channel in future experiments. We also see that the contributions are principally from $l \neq 0$ channels, indicating the importance of anisotropic interactions. We note that our analysis is restricted to two-body inelastic processes only, and this restriction may contribute to the residual quantitative deviations between 
the experimental and theoretical spectra. 

\section{Summary and Conclusions}

We observe a strong feature in the inelastic rate of the $^{174}$Yb $^3P_2,m_J=-1$ + $^6$Li $^2S_{1/2},|m_J,m_I\rangle=|-1/2,+1\rangle$ system at $450\,$G. When combined with a theoretical model of the system which fits the obtained inelastic rate spectrum, this provides evidence for anisotropy-induced magnetic Feshbach resonances in this collisional system. Future work includes a more careful study of the nature of these Li-Yb* resonances and their applicability towards molecule formation. A direct ${^1S_0} \rightarrow {^3P_2(m)}$ optical transfer scheme \cite{Taka507} will allow experiments to efficiently access each Zeeman state individually. Of particular interest is the lowest Zeeman state of fermionic $^{173}$Yb where the intra-species inelastic effects should be suppressed due to Pauli-blocking of $s-$wave collisions and background inter-species inelastic effects should be reduced from being in the lowest energy collision channel in the ${^3P_2}$ Zeeman manifold.

We thank Alaina Green for experimental assistance. We gratefully acknowledge financial support from NSF Grants No. PHY-1306647, No. PHY-1308573, AFOSR Grants No. FA9550-12-10051, No. FA9550-14-1-0321, and ARO MURI Grant No. W911NF-12-1-0476.


\begin{thebibliography}{40}
\bibitem{SIM1} A. Micheli, G.K. Brennen, and P. Zoller, Nature Physics \textbf{2}, 341 (2006).
\bibitem{FUND1} J.J. Hudson, D.M. Kara, I.J. Smallman, B.E. Sauer, M.R. Tarbutt, and E.A. Hinds, Nature \textbf{473}, 493 (2011).
\bibitem{FESH1} S. Inouye, M.R. Andrews, J. Stenger, H.-J. Miesner, D.M. Stamper-Kurn, W. Ketterle, Nature \textbf{392}, 151 (1998).
\bibitem{FESH2} Ph. Courteille, R.S. Freeland, D.J. Heinzen, F.A. van Abeelen, and B.J. Verhaar, Phys. Rev. Lett. \textbf{81}, 69 (1998).
\bibitem{FESH3} V. Vuleti{\'c}, A.J. Kerman, C. Chin, and S. Chu, Phys. Rev. Lett. \textbf{82}, 1406 (1999).
\bibitem{Cs2} J.G. Danzl, E. Haller, M. Gustavsson, M.J. Mark, R. Hart, N. Bouloufa, O. Dulieu, H. Ritsch, H-C. Nagerl, Science \textbf{321}, 1062 (2008).
\bibitem{KRb} K.K. Ni, S. Ospelkaus, M.H.G. de Miranda, A. Pe'er, B. Neyenhuis, J.J. Zirbel, S. Kotochigova, P.S. Julienne, D.S. Jin, J. Ye, Science \textbf{322}, 231 (2008).
\bibitem{YBLI1} V. Ivanov, A. Khromov, A. Hansen, W. Dowd, F. M{\"u}nchow, A. Jamison, and S. Gupta, Phys. Rev. Lett. \textbf{106}, 153201 (2011).
\bibitem{YBLI2} H. Hara, Y. Takasu, Y. Yamaoka, J. Doyle, Y. Takahashi, Phys. Rev. Lett. \textbf{106}, 205304 (2011).
\bibitem{RBSR1} B. Pasquiou, A. Bayerle, S.M. Tzanova, S. Stellmer, J. Szczepkowski, M. Parigger, R. Grimm, and F. Schreck, Phys. Rev. A. \textbf{88}, 023601 (2013).
\bibitem{RBYB1} F. Baumer, F. M{\"u}nchow, A. G{\"o}rlitz, S.E. Maxwell, P.S. Julienne, and E. Tiesinga, Phys. Rev. A \textbf{83}, 040702 (2011).
\bibitem{NEWFESH} P.S. Zuchowski, J. Aldegunde, and J.M. Hutson, Phys. Rev. Lett. \textbf{105}, 153201 (2010).
\bibitem{CALC1} D. Hansen, J. Mohr, and A. Hemmerich, Phys. Rev. A \textbf{67}, 021401 (2003).
\bibitem{SR1} S.B. Nagel, C.E. Simien, S. Laha, P. Gupta, V.S. Ashoka, and T.C. Killian, Phys. Rev. A \textbf{67}, 011401 (2003).
\bibitem{Yb1} A. Yamaguchi, S. Uetake, D. Hashimoto, J.M. Doyle, and Y. Takahashi, Phys. Rev. Lett. \textbf{101}, 233002 (2008).
\bibitem{YBMETAFESH} S. Kato, S. Sugawa, K. Shibata, R. Yamamoto, and Y. Takahashi, Phys. Rev. Lett. \textbf{110}, 173201 (2013).
\bibitem{METAFESH} A. Petrov, E. Tiesinga, and S. Kotochigova, Phys. Rev. Lett. \textbf{109}, 103002 (2012).
\bibitem{LIYBMETA} A. Khramov, A. Hansen, W. Dowd, R. Roy, C. Makrides, A. Petrov, S. Kotochigova, and S. Gupta, Phys. Rev. Lett. \textbf{112}, 033201 (2014).
\bibitem{LIYBMETAFESH} M.L. Gonz{\'a}lez-Mart{\'i}nez and J.M. Hutson, Phys. Rev. A \textbf{88}, 020701 (2013).
\bibitem{PRODUCTION} A. Hansen, A. Khramov, W. Dowd, A. Jamison, B. Plotkin-Swing, R. Roy, and S. Gupta, Phys. Rev. A. \textbf{87}, 013615 (2013).
\bibitem{BOWERS} C.J. Bowers, D. Budker, S.J. Freedman, G. Gwinner, J.E. Stalnaker, and D. DeMille, Phys. Rev. A \textbf{59} 3513 (1999).
\bibitem{footspinpure} The choice to use the $^3P_2, m_J=-1$ state is primarily a technical one. The ideal experiment would use the $m_J=-2$ substate to reduce the number of decay channels for the system. However, our optical pumping scheme will always produce $m_J=-1$ together with $m_J=-2$. At our optical trap wavelength, there is no relative orientation of electric and magnetic field orientation that produces a sufficiently weaker trap for $m_J=-1$ to facilitate purification via spilling. Transfer of our purified sample to the $m_J=-2$ substate via a radiofrequency pulse would require the polarization of the optical trap to also be rapidly changed to provide sufficient trapping, a technique that could be implemented in the future.
\bibitem{foottemp} Including these temperature variations in the analysis led to $<\,5\%$ variation in the extracted inelastic coefficients, which is smaller than the reported error bars.
\bibitem{footoverlap} The interspecies overlap during the mixture evolution time can be affected by relative cloud center-of-mass motion from small intensity imbalances in the two counter-propagating $404\,$nm excitation beams, which we experimentally minimize. Another contributor is a small magnetic field gradient along the weakest trap axis which differentially affect the two species.
\bibitem{JulienneUnitarity} Z. Idziaszek and P. Julienne, Phys. Rev. Lett. \textbf{104}, 113202 (2010)
\bibitem{SvetTheory} A. Petrov, C. Makrides, S. Kotochigova, arXiv:1502.04973 (2015).
\bibitem{Taka507} S. Uetake, R. Murakami, J. Doyle, and Y. Takahashi, Phys. Rev. A \textbf{86}, 032712 (2012).

\end{thebibliography}
\end{document}